\documentclass[aps,pra,twocolumn,amsmath,amssymb,nofootinbib,showpacs]{revtex4}
\usepackage[english]{babel}
\usepackage{latexsym}
\usepackage{graphics}
\usepackage{epsfig}
\usepackage{color}
\usepackage[latin1]{inputenc}
\usepackage[T1]{fontenc}

\newcommand{\epsfboxmod}[1]{\epsfbox{#1.eps}}
\newcommand{\infig}[2]{\begin{center}
                                    \mbox{ \epsfxsize #1 \epsfboxmod{#2}}
                                    \end{center}}

\newcommand{\ie}{{i.e.}}

\newcommand{\heaviside}{\Theta}

\newcommand{\av}[1]{\langle #1 \rangle}
\newcommand{\li}{\textrm{Li}_{2}}

\newcommand{\Vr}{V_\textrm{\tiny R}}
\newcommand{\sigmar}{\sigma_\textrm{\tiny R}}
\newcommand{\epsilonr}{\epsilon_\textrm{\tiny R}}
\newcommand{\ksuppr}{k_\textrm{\tiny c}}

\newcommand{\corr}[1]{C_{#1}}
\newcommand{\redcorr}[1]{c_{#1}}
\newcommand{\hatredcorr}[1]{\hat{c}_{#1}}

\newcommand{\redcorrA}{c_a}
\newcommand{\hatredcorrA}{\hat{c}_a}

\newcommand{\lyap}{\gamma}

\newcommand{\dlyap}[1]{\gamma^{(#1)}}

\newcommand{\fn}[1]{f_{#1}}

\newcommand{\refeq}[1]{(\ref{#1})}

\begin{document}

\title{One-dimensional Anderson localization in certain correlated random potentials}

\author{P.~Lugan, A.~Aspect, and L.~Sanchez-Palencia}
\affiliation{
Laboratoire Charles Fabry de l'Institut d'Optique,
CNRS and Univ.~Paris-Sud,
Campus Polytechnique, 
RD 128, 
F-91127 Palaiseau cedex, France}

\author{D.~Delande$^{1}$, B.~Gr\'emaud$^{1,2}$, C.A.~M\"uller$^{1,3}$, and C.~Miniatura$^{2,4}$}
\affiliation{$^{1}$Laboratoire Kastler-Brossel, UPMC, ENS, CNRS;
4 Place Jussieu, F-75005 Paris, France \\
$^{2}$Centre for Quantum Technologies, National University of Singapore, 3 Science Drive 2, Singapore 117543, Singapore \\
$^{3}$Physikalisches Institut, Universit\"at Bayreuth, D-95440 Bayreuth, Germany \\
$^{4}$Institut Non Lin\'eaire de Nice, UNS, CNRS; 1361 route des Lucioles, F-06560 Valbonne}

\date{\today}

\begin{abstract}
We study Anderson localization of ultracold atoms
in weak, one-dimensional speckle potentials,
using perturbation theory beyond Born approximation.
We show the existence of a series of sharp crossovers (effective mobility edges)
between energy regions where localization lengths differ by orders of magnitude.
We also point out that the correction to the Born term explicitly depends
on the sign of the potential.
Our results are in agreement with numerical calculations in a regime relevant for experiments.
Finally, we analyze our findings in the light of a diagrammatic approach.
\end{abstract}


\maketitle

\section{Introduction}
\label{intro}
Anderson localization (AL) of single electron wave functions~\cite{anderson1958}, first proposed to understand certain metal-insulator transitions, is now considered an ubiquitous phenomenon,
which can happen for any kind of waves propagating in a medium with random impurities~\cite{akkermans2006,vantiggelen1999}. It can be understood as a coherent interference effect of waves multiply scattered from random defects, yielding localized waves with exponential profile,  and resulting in complete suppression of the usual diffusive transport associated with incoherent wave scattering~\cite{lee1985}.
So far, AL has been reported for
light waves in diffusive media~\cite{wiersma1997,storzer2006}
and photonic crystals~\cite{schwartz2007,lahini2008},
sound waves~\cite{hu2008},
or microwaves~\cite{chabanov2000}.
Ultracold atoms have allowed
studies of AL in  momentum space~\cite{raizen,chabe2008}
and recently
direct observation of localized atomic matter waves~\cite{billy2008,roati2008}.

In one-dimensional (1D) systems,
all states are localized, and the localization length is simply proportional
to the transport mean-free path~\cite{beenakker1997}.
However, this strong property should not hide that
long-range correlations can induce subtle effects in 1D models of disorder,
in particular those whose power spectrum has a finite support~\cite{izrailev1999,lsp2007}.
Examples are random potentials resulting from laser speckle
and used in experiments with ultracold atoms~\cite{billy2008,clement2006,fallani2008}.
Indeed, by construction~\cite{goodman2007}, speckles
have no Fourier component beyond a certain value $2\ksuppr$,
and the Born approximation
predicts no back-scattering and no localization for
{atoms with momentum $\hbar k>\hbar\ksuppr$}.
This defines an {\it effective mobility edge} at $k=\ksuppr$~\cite{lsp2007},
clear evidence of which has been reported~\cite{billy2008}.

Beyond this analysis --relevant for systems of moderate size \cite{lsp2007,billy2008}--
study of AL in correlated potentials beyond the effective mobility edge
requires more elaborated approaches.
{In Ref.~\cite{tessieri2002},} disorder with symmetric probability distribution was
studied, and examples were exhibited, for which exponential localization occurs even
for $k>\ksuppr$ although with a much longer localization length than for $k<\ksuppr$.
It was also concluded that for Gaussian disorder, there is a second
effective mobility edge at $2\ksuppr$, while for non-Gaussian disorder, it is generally not so.
These results do not apply to speckle potentials
{whose} probability distribution is \emph{asymmetric}.
{Moreover,
although speckle potentials are not Gaussian, they
derive from the squared modulus of a Gaussian field,
and, as we will show, the conclusions of Ref.~\cite{tessieri2002} must be re-examined.
Hence, considering speckle potentials
presents a twofold interest.
First, they form an original class of non-Gaussian disorder which can inherit properties of
an underlying Gaussian process.
Second, they are easily implemented in experiments with ultracold atoms
where the localization length can be directly measured~\cite{billy2008}.
}

\begin{figure}[b!]
\begin{center}
\infig{19.em}{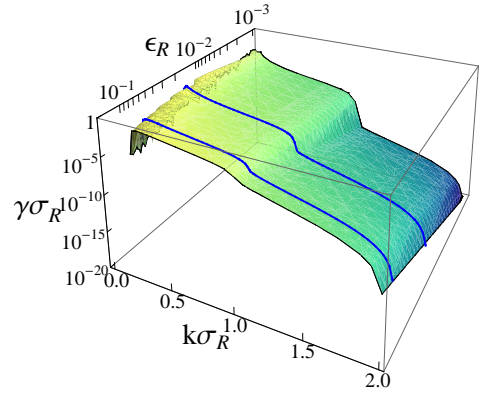}
\end{center}
\vspace{-0.5cm}
\caption{\small
(Color online)
Lyapunov exponent $\gamma$ calculated two orders beyond the Born approximation
for particles in 1D speckle potentials created with a square diffusive plate,
versus the particle momentum {$\hbar k$} and the strength of disorder $\epsilonr=2m\sigmar^2\Vr/\hbar^2$
($\Vr$ and $\sigmar$ are the amplitude and correlation length of the disorder).
The {solid blue} lines correspond to $\epsilonr=0.1$ and $\epsilonr=0.02$.
} 
\label{fig:gamma3D}
\end{figure}

In this work, we study AL in speckle potentials beyond the Born approximation,
using perturbation theory~\cite{gurevich2009},
numerical calculations,
and diagrammatic methods.
We find that
there exist several effective mobility edges at
$\ksuppr^{(p)}=p\ksuppr$ with {integer} $p$,
such that AL in the successive intervals $\ksuppr^{(p-1)}<k<\ksuppr^{(p)}$
results from scattering processes of increasing order.
Effective mobility edges are thus characterized by sharp crossovers
in the $k$ dependence of the Lyapunov exponent (see Fig.~\ref{fig:gamma3D}).
We prove this for the first two effective mobility edges by calculating
the three lowest-order terms, and
give general arguments for any $p$.
In addition, we discuss the effect of odd terms that appear in the Born series due to
the asymmetric probability distribution of speckle potentials.

\section{Speckle potentials}
\label{sec:phaseform.speckle}
Let us first recall the main properties of speckle potentials.
Optical speckle is obtained by transmission of a laser beam through a
medium with a random phase profile, such as a ground glass plate~\cite{goodman2007}.
The resulting complex electric field $\mathcal{E}$ {is} a sum of
independent random variables {and forms} a
Gaussian process. In such a light field, atoms experience a random potential
proportional to {the intensity} $|\mathcal{E}|^2$.
Defining the zero of energies so that $\av{V}=0$, the random potential is thus
\begin{equation}
V(z)=\Vr \times \left(|a(z/\sigmar)|^2 - \av{|a(z/\sigmar)|^2}\right)
\label{eq:VrVSa}
\end{equation}
where the quantities $a(u)$ are complex Gaussian variables proportional to the electric field
$\mathcal{E}$,
and
$\sigmar$ and $\Vr$ feature characteristic length and strength scales of the
random potential
(The precise definition of $\Vr$ and $\sigmar$ may depend on the model of disorder; see below).
{In contrast}, $V(z)$ is not a Gaussian variable and
its probability distribution is a decaying exponential, \ie\ asymmetric.
The sign of $\Vr$ is thus relevant and
can be either positive or negative for "blue"- and "red"-detuned laser light respectively.
However, the random potential $V(z)$ inherits properties of the
underlying Gaussian field $a(u)$.
For instance, all potential correlators $\redcorr{n}$ are
completely determined by the field correlator $\redcorrA(u)=\av{a(0)^*a(u)}$
via
\begin{equation}
\av{a_1^* ... a_p^* \times a_{1} ... a_{p}}
=
\sum_{\Pi} \av{a_1^*a_{\Pi(1)}} ... \av{a_p^*a_{\Pi(p)}},
\label{eq:wick}
\end{equation}
where $a_{p'}=a(z_{p'}/\sigmar)$
and $\Pi$ describes {the} $p!$ permutations of $\{1, ..., p\}$.
Hence, $\redcorr{2}(u)=|\redcorrA (u)|^2$ and
defining $a(u)$ so that $\av{|a(u)|^2}=1$,
we have $\sqrt{\av{V(z)^2}}=|\Vr|$.
{Also}, since speckle results from interference between light waves of wavelength $\lambda_\textrm{L}$
coming from a finite-size aperture of angular width $2\alpha$,
the Fourier transform of the field correlator has no component beyond
$\ksuppr=2\pi\sin \alpha /\lambda_\textrm{L}$,
and
$\redcorrA$ has always a finite support:
\begin{equation}
\hatredcorrA(q)=0 ~~~~~\textrm{for}~|q|>\ksuppr\sigmar \equiv 1.
\label{eq:support}
\end{equation}
As a consequence, the Fourier transform of the potential correlator also has a finite support:
$\hatredcorr{2}(q)=0$ for $|q|>2$.

\section{Phase formalism}
\label{sec:phaseform}
Consider now a particle of energy $E$ in a 1D random potential $V(z)$ with zero 
statistical average [$V(z)$ need not be a speckle potential here].
The particle wave function $\phi$ can be written in phase-amplitude representation
\begin{equation}
\phi(z) = r(z) \sin \left[\theta(z)\right];~~~~~
\partial_z \phi = k r(z) \cos \left[\theta(z)\right],
\label{eq:phasefunct}
\end{equation}
which proves convenient to capture the asymptotic decay of the wave function
(here $k=\sqrt{2mE/\hbar^2}$ is the particle wave vector in the absence of disorder).
It is easily checked that the Schrödinger equation is then equivalent to
the coupled equations
\begin{eqnarray}
&& \partial_z \theta (z) = k\left[1 - \left(V(z)/E\right) \sin^2 \left(\theta (z)\right)\right]
   \label{eq:phaseEq1} \\
&& \ln [r(z)/r(0)] = k \int_0^z \textrm{d}z' \left(V(z')/2E\right) \sin \left(2\theta (z')\right).
   \label{eq:phaseEq2}
\end{eqnarray}
Since Eq.~(\ref{eq:phaseEq1}) is a closed equation for the phase $\theta$, it is
straightforward to develop the perturbation series of $\theta$ in increasing powers of
$V$. Reintroducing the solutions at different orders into Eq.~(\ref{eq:phaseEq2})
yields the corresponding series for the amplitude $r (z)$ and the Lyapunov exponent:
\begin{equation}
\lyap (k) = \lim_{|z| \to \infty}\frac{\av{\ln [r(z)]}}{|z|} = \sum_{n \ge 2} \dlyap{n} (k).
\label{eq:pertlyap}
\end{equation}
The $n$th-order term $\dlyap{n}$ is thus expressed as a function of the
$n$-point correlator
$\corr{n}(z_{1},...,z_{n-1})=\av{V(0)V(z_{1})...V(z_{n-1})}$
of the random potential, which we write
$\corr{n}(z_{1},...,z_{n-1})= \Vr^n \redcorr{n}\left({z_{1}}/{\sigmar},...,{z_{n-1}}/{\sigmar}\right)$.
Up order $n=4$, we find
\begin{equation}
\dlyap{n} = \sigmar^{-1} \left(\frac{\epsilonr}{k\sigmar}\right)^n \fn{n}(k\sigmar)
\label{eq:dlyap}
\end{equation}
where $\epsilonr = {2 m \sigmar^2 \Vr}/{\hbar^2}$ and
\begin{eqnarray}
\fn{2}(\kappa) &=& +\frac{1}{4}
	\int_{-\infty}^{0}\textrm{d}u\
	\redcorr{2}(u)\cos(2\kappa u)
\label{eq:lyap1} \\
\fn{3}(\kappa) &=&  -\frac{1}{4}
	\int_{-\infty}^{0} \textrm{d}u
	\int_{-\infty}^{u} \textrm{d}v\ \redcorr{3}(u, \! v)\sin(2\kappa v)
\label{eq:lyap2} \\
\fn{4}(\kappa) &=&  -\frac{1}{8}
	\int_{-\infty}^{0} \!\! \textrm{d}u
	\int_{-\infty}^{u} \!\! \textrm{d}v
	\int_{-\infty}^{v} \!\! \textrm{d}w\ \redcorr{4}(u, v, w) \label{eq:lyap3} \\
&& \hspace{0.8cm} \times \big\{ \! 2\cos(2\kappa w) \! + \! \cos[2\kappa(v \! + \! w \! - \! u)] \! \big\}. \nonumber
\end{eqnarray}
Note that the compact form~(\ref{eq:lyap3}) is valid provided that
oscillating terms, which may appear from terms in $c_4$ that can be factorized
as $c_2$ correlators, are appropriately regularized at infinity.
Note also that in Eq.~(\ref{eq:dlyap}), the coefficients $(\epsilonr/k\sigmar)^n$ diverge
for $k\to 0$, while the exact $\lyap(k)$ remains finite for any
$\epsilonr$~\cite{derrida1984}.
This signals a well-known breakdown of the perturbative {approach}.
Conversely, the perturbative expansion is valid when $\gamma(k) \ll k$
(for $k\to 0$),
\ie\ when the localization length
exceeds the particle wavelength,
a physically satisfactory criterion.

\section{One-dimensional Anderson localization in speckle potentials}
\label{sec:AL}
\subsection{Analytic results}
Let us now examine the consequences of the peculiar properties of
speckle potentials in the light of the above {perturbative} approach.
For clarity, we restrict ourselves to 1D speckle potentials
created by square diffusive plates as in Refs.~\cite{billy2008,clement2006} for which
$\redcorrA(u)=\sin (u)/u$
and $\hatredcorrA(q) \propto \heaviside (1-|q|)$
where $\Theta$ is the Heaviside step function~\cite{noteGeneralization}.
Using
Eqs.~(\ref{eq:lyap1}) and (\ref{eq:lyap2}), we find
\begin{eqnarray}
\fn{2}(\kappa) &=& \frac{\pi}{8}\Theta(1-\kappa)(1-\kappa) \label{eq:sq:f2a} \\
\fn{3}(\kappa) &=& -\frac{\pi}{4}\Theta(1-\kappa)\left[
		(1 \! - \! \kappa)\ln\left({1 \! - \! \kappa}\right)+
		\kappa\ln\left({\kappa}\right)
		\right]~~~~~ \label{eq:sq:f2b}
\end{eqnarray}

\begin{figure}[t!]
\begin{center}
\infig{24.em}{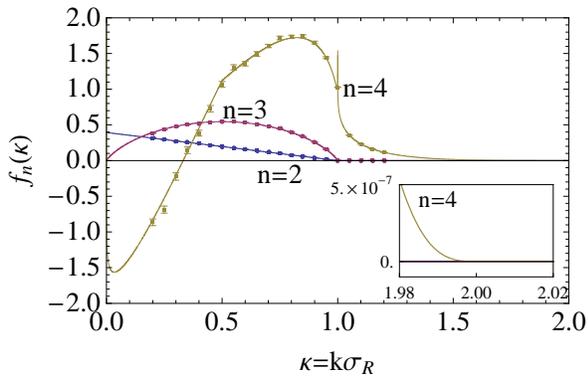}
\end{center}
\caption{
(Color online)
Functions $\fn{n}$ for $n=2$, $3$ and $4$ for a
speckle potential created with a square diffusive plate 
[solid lines; see Eqs.~(\ref{eq:sq:f2a}) and (\ref{eq:sq:f2b}) and the Appendix]
and comparison with numerical calculations (points with error bars).
The inset is a magnification of function $\fn{4}$ around $\kappa=2$.}
\label{fig:fs}
\end{figure}

The functions $\fn{2}$ and $\fn{3}$ are simple and vanish for $\kappa \geq 1$
(see Fig.~\ref{fig:fs}).
This property is responsible for the existence of the first effective mobility edge
at $k=\ksuppr$~\cite{lsp2007},
such that
$\lyap(k)\sigmar \sim (\epsilonr/k\sigmar)^2$ for $k \lesssim \sigmar^{-1}$
while
$\lyap(k)\sigmar = \textrm{O}(\epsilonr/k\sigmar)^4$ for $k \gtrsim \sigmar^{-1}$.
{The fact that $\fn{3}$ vanishes in the same interval ($\kappa \geq 1$) as $\fn{2}$
exemplifies the general property that
odd-$n$ terms cannot be leading terms in any range of $k$
because $\lyap (k)$ must be positive whatever the sign of $\Vr$.
For $\kappa<1$ however,} $\fn{3}(\kappa)$ is not identically zero
owing to the asymmetric
probability distribution in speckle potentials.
The term $\dlyap{3}$ 
can thus be either
positive or negative depending on the sign of $\Vr$~\cite{gurevich2009}.

The function $\fn{4}$ is found similarly from Eq.~(\ref{eq:lyap3}).
While its expression is quite 
complicated (see the Appendix), its
behavior is clear when plotted (see Fig.~\ref{fig:fs}).
Let us emphasize some {of its} important features.
First, there is a discontinuity of the derivative of $\fn{4}$ at $\kappa = 1/2$.
Second, we find a very narrow logarithmic divergence,
$\fn{4}(\kappa) \sim -(\pi/32)\ln |1-\kappa|$  at $\kappa = 1$,
which signals a singularity of the perturbative {approach}
(note that it does not appear in Fig.~\ref{fig:gamma3D}
due to finite resolution {of the plot}).
Finally, the value $\kappa = 2$ corresponds to the boundary of the support of $\fn{4}$,
showing explicitly the existence of a second effective mobility edge at $k=2\sigmar^{-1}$.
{Hence, while $\lyap(k)\sigmar \sim (\epsilonr/k\sigmar)^4$ for $k \lesssim 2\sigmar^{-1}$,
we have $\lyap(k)\sigmar = \textrm{O}(\epsilonr/k\sigmar)^6$ for $k \gtrsim 2\sigmar^{-1}$,
since $\fn{4}(\kappa)$ as well as $\fn{5}(\kappa)$ vanish for $\kappa\geq 2$.}

\subsection{Numerics}
\begin{figure}[t!]
\begin{center}
\vspace{0.23cm}
\infig{23em}{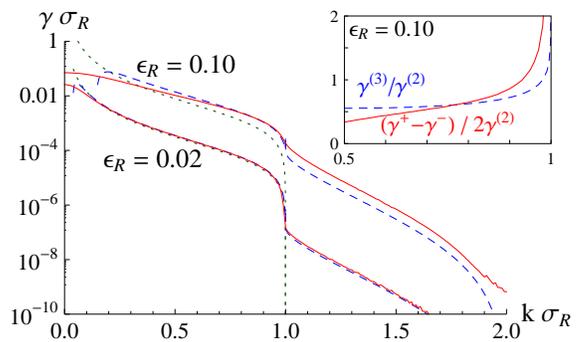}
\end{center}
\caption{
(Color online)
Lyapunov exponent $\lyap(k)$ versus the particle momentum $k$
as determined numerically (solid red lines)
and by perturbation theory up to order $4$ (dashed blue lines),
for a speckle potential created with a square plate.
The dotted green lines are the Born term.
Inset: comparison of odd and even contributions 
in the Born series for $\epsilonr=0.1$.
}
\label{fig:gamma}
\end{figure}

In order to test the validity of the perturbative approach for 
experimentally relevant parameters, we have performed 
numerical calculations using a transfer matrix approach.
The results are plotted in Fig.~\ref{fig:gamma}:
$\epsilonr=0.02$ corresponds to
$\Vr/\hbar=2\pi\times 16$Hz in Fig.~3 of Ref.~\cite{billy2008}
and $\epsilonr=0.1$ to $\Vr/\hbar=2\pi\times 80$Hz in 
Fig.~3 and to Fig.~4 of Ref.~\cite{billy2008}.
For $\epsilonr=0.02$, the agreement between analytical and 
numerical results is excellent.
The effective mobility edge at $k=\sigmar^{-1}$ is very clear:
we find a sharp step for $\lyap(k)$ of about 2 orders of magnitude.
For $\epsilonr=0.1$, we find the same trend
but with a smoother and smaller step (about one order of magnitude).
In this case, although the Born term for $k \lesssim \sigmar^{-1}$
and the fourth-order term for $k \gtrsim \sigmar^{-1}$ provide 
reasonable estimates (within a factor of $2$),
higher-order terms --which may depend on the sign of $\Vr$--
contribute significantly.

The contribution of the odd terms
can be extracted by taking $\gamma^{+}-\gamma^{-}$,
where $\gamma^{\pm}$ are the Lyapunov exponents obtained
for positive and negative disorder amplitude of same modulus
$|\Vr|$, respectively.
As shown in the inset of Fig.~\ref{fig:gamma},
the odd terms {range from $30\%$ to $70\%$} of
the Born term
for $0.6 \lesssim k\sigmar \lesssim 0.9$ and $\epsilonr=0.1$,
and are {of the order of} $\dlyap{3}$ in weak disorder
and away from the divergence at $k=\sigmar^{-1}$.
This shows that the first correction $\dlyap{3}$ to the Born term
can be relevant in {experiments}.

For completeness,
{we have calculated the $\fn{n}(\kappa)$
as the coefficients of fits in powers of $\epsilonr/k\sigmar$
using series of calculations of $\lyap (k)$
at fixed $k$ and various $\epsilonr$}.
As shown in Fig.~\ref{fig:fs}, the agreement
with
the analytic formulas is excellent.
In particular, the numerics
reproduce the predicted
kink at $\kappa = 1/2$.
The logarithmic singularity around
$\kappa=1$ being very narrow, we did not
attempt to study it.

\section{Diagrammatic analysis}
\label{diagrams}
\newcommand{\fieldcorrFIG }{\raisebox{0ex}{\includegraphics{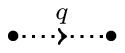}}}
\newcommand{\potcorrFIG }{\raisebox{0ex}{\includegraphics{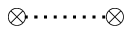}}}
\newcommand{\UonewithkFIG }{\raisebox{-3.5ex}{\includegraphics{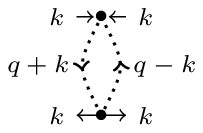}}}
\newcommand{\UtwowithkFIG }{\raisebox{-3.5ex}{\includegraphics{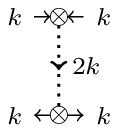}}}
\newcommand{\UthreewithkFIG }{\raisebox{-3.5ex}{\includegraphics{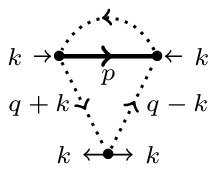}}}

Let us finally complete our analysis using diagrammatic methods,
which allow us to exhibit momentum exchange in scattering processes as compact graphics,
and thus to identify effective mobility edges in a quite general way.
In 1D, the localization length
can be calculated from the backscattering probability of
$\av{|\psi|^2}$ using quantum transport theory.
The irreducible diagrams of elementary scattering processes
in speckle potentials have been identified in Ref.~\cite{kuhn2007}.

To lowest order in $\epsilonr$ (Born approximation),
the average intensity of a plane wave with wave vector $k$
backscattered by the random potential is described by 
\begin{equation}
\label{eq:U12k}
U_{2} (k) = \UonewithkFIG =: \UtwowithkFIG.
\end{equation}
The upper part of the diagram represents $\psi$ (particle) and the lower part its conjugate $\psi^*$ (hole).
The dotted line $\fieldcorrFIG =\epsilonr \hat{c}_a(q)$ represents the field correlator;
simple closed loops over field correlations can be written as a potential correlation $\potcorrFIG$.
Backscattering requires diagram \refeq{eq:U12k} to channel a momentum $2k$,
entering at the particle, down along the potential correlations to the hole.
Therefore, the diagram vanishes for
$k\sigmar>1$.

At order $\epsilonr^3$, the only possible contribution is
\begin{equation}\label{eq:U3withk}  
U_{3}(k) = \ \UthreewithkFIG \ + c.c. 
\end{equation}
The straight black line stands for the particle propagator 
$[E_k-E_p+i0]^{-1}$ at intermediate momentum $p$.
Diagram~\refeq{eq:U3withk} features two vertical field correlation lines,
just as diagram~\refeq{eq:U12k}, and thus vanishes at the same threshold 
$k=\sigmar^{-1}$.
Evaluating two-loop diagram~(\ref{eq:U3withk}), we recover precisely
contribution~\refeq{eq:sq:f2b}.

\begin{figure}[t!]
\begin{center}
\includegraphics[width=0.8\linewidth]{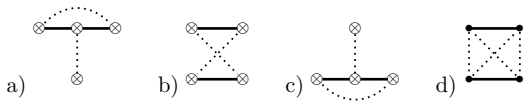}
\end{center}
\vspace{-1em}
\caption{Relevant fourth-order backscattering contributions.
Contrary to the case of uncorrelated potentials~\cite{Gornyi2007,Cassam1994},
the sum of diagrams (a)-(c) \emph{does not} give zero
for speckle potentials; only diagrams (b) and (d) contribute for
$k\sigmar\in[1,2]$.}
\label{fig:diag4}
\end{figure}

Many diagrams contribute to order $\epsilonr^4$. 
First there are the usual backscattering contributions with pure intensity correlations
[Fig.~\ref{fig:diag4}(a)-\ref{fig:diag4}(c)].
Both Figs.~\ref{fig:diag4}(a) and \ref{fig:diag4}(c) have a single vertical intensity correlation and vanish for $k>\sigmar^{-1}$.
In contrast,  the crossed diagram [\ref{fig:diag4}(b)] has two vertical intensity correlation lines and can thus accommodate momenta up to $k=2\sigmar^{-1}$.
Performing the integration, we find that this diagram reproduces
those contributions
to $\fn{4}(\kappa)$ for $\kappa \in [1,2]$ that contain factorized correlators
(see the Appendix).
Second, there are nine more diagrams, all with non-factorizable field correlations~\cite{kuhn2007}.
A single one has not two, but four vertical field correlation lines, shown in  Fig.~\ref{fig:diag4}(d), 
and contributes for $k\sigmar\in[1,2]$.
Carrying out the three-loop integration, we recover exactly the non-factorizable contributions
to $\fn{4}(\kappa)$ for $\kappa \in [1,2]$.

\section{Conclusion}
\label{conclusion}
We have developed {perturbative} and diagrammatic approaches {beyond the Born approximation},
suitable {to study} 1D AL in correlated disorder
with possibly 
asymmetric probability distribution.
In speckle potentials,
the $k$ dependence of the Lyapunov exponent exhibits sharp crossovers
(effective mobility edges) separating regions
where AL is due to scattering processes of increasing order.
We have shown it explicitly
for $k=\sigmar^{-1}$ and $k=2\sigmar^{-1}$, and
we infer that there is a series of effective mobility edges at $k=p\sigmar^{-1}$
with {integer} $p$ since, generically,
diagrams with $2p$
field correlations or $p$ intensity correlations
can contribute up to $k=p\sigmar^{-1}$.
This is because, although
speckles are not Gaussian, they 
{derive from} a Gaussian field.
Finally, exact numerics support
our analysis for experimentally relevant parameters,
{and indicate the necessity to use higher-order terms in the Born series,
even for $k<\sigmar^{-1}$}.
Hence, important features
that we have pointed out,
such as odd terms in the Born series {for $k<\sigmar^{-1}$} and exponential localization
for $k>\sigmar^{-1}$,
{should be} observable {experimentally}.

\section*{ACKNOWLEDGMENTS}
Stimulating discussions with P.~Bouyer, V.~Josse, T.~Giamarchi and B.~van Tiggelen
are acknowledged.
This research was supported by the French
CNRS,
ANR,
MENRT,
Triangle de la Physique and
IFRAF.

\begin{appendix}
\section*{APPENDIX}

\noindent
Here, we give the explicit formula of the function $\fn{4}(\kappa)$ for a speckle potential created by a square diffusive plate,
such that the fourth-order term in the Born expansion of the
Lyapunov exponent $\lyap$ reads
$\dlyap{4} = \sigmar^{-1} \left(\frac{\epsilonr}{k\sigmar}\right)^{4} \fn{4}(k\sigmar)$.
The function $\fn{4}(\kappa)$ is the sum of three terms with different supports,
\begin{equation}
\fn{4}(\kappa) = \fn{4}^{[0,1/2]}(\kappa) + \fn{4}^{[0,1]}(\kappa) + \fn{4}^{[1,2]}(\kappa),
\nonumber
\end{equation}
where $\fn{4}^{[\alpha,\beta]}(\kappa)$ lives on the interval $\kappa \in [\alpha,\beta]$,
and
\begin{widetext}
\begin{eqnarray}
\fn{4}^{[0,{1}/{2}]}(\kappa)
	&=&
-\frac{\pi^3}{16}(1-2\kappa)
\nonumber\\
\fn{4}^{[0,1]}(\kappa)
	&=&
\frac{\pi}{64}\left\{
4-6\kappa
-\frac{10\pi^2}{3}(1-2\kappa)
-(4-2\kappa)\ln(\kappa) \right.
-\left(\frac{5}{\kappa}-3\kappa\right)\ln(1-\kappa)
+\left(\frac{1}{\kappa}+\kappa\right)\ln(1+\kappa)
	\nonumber\\
	&&
-(4-8\kappa)\ln^2(\kappa)
+22(1-\kappa)\ln^2(1-\kappa)
+(18+14\kappa)\ln^2(1+\kappa)
	\nonumber\\
	&&
-16(1-\kappa)\ln(1-\kappa)\ln(\kappa)
-4(1-\kappa)\ln(1-\kappa)\ln(1+\kappa)
-32(1+\kappa)\ln(\kappa)\ln(1+\kappa)
	\nonumber\\
	&&
-24(1+\kappa)\li(\kappa)
+32(1+\kappa)\li\left(\frac{\kappa}{1+\kappa}\right)
\left.
-8\kappa\li\left(\frac{2\kappa}{1+\kappa}\right)
-8(1-2\kappa)\li\left(2-\frac{1}{\kappa}\right)
\right\}
	\nonumber\\
\fn{4}^{[1,2]}(\kappa)
	&=&
\frac{\pi}{32}\bigg\{-2+\left(1+\frac{\pi^2}{3}\right)\kappa
+ 4\kappa \li(1-\kappa)
-\left(\frac{2}{\kappa}-2+\kappa\right)\ln(\kappa-1)
-2(\kappa-1)\ln^2(\kappa-1)+4\kappa \ln(\kappa-1)\ln(\kappa) \bigg\} \nonumber
\end{eqnarray}
\end{widetext}
where
$\li (z) = \int_{z}^{0}\mathrm{d}t\ {\ln(1-t)}/{t} = \sum_{k=1}^{\infty}{z^k}/{k^2}$
is the dilogarithm function.

\end{appendix}



\begin{thebibliography}{99}
\bibitem{noteGeneralization} 
Our results can be extended to any 1D random potential
that fulfill conditions~(\ref{eq:VrVSa})-(\ref{eq:support}) with
similar conclusions.
\end{thebibliography}
\end{document}